\documentclass{PoS}

\newcommand{\sor}{4U~1722--30}
\newcommand{\intg}{\emph{INTEGRAL}}
\newcommand{\xte}{\emph{RXTE}}

\title{High energy characteristics of Neutron star low mass X-ray binaries as seen by INTEGRAL: the \sor\/ case}

\ShortTitle{The \sor\/ case}

\author{\speaker{Antonella Tarana}\thanks{This research has made use of data obtained with INTEGRAL an international collaboration. This work has been supported by the Italian Space Agency through the grant ASI/INAF I/0033/10/0. The authors thanks M. Federici for the maintenance of \intg\/ data archive and software for the AVES cluster. A.T. acknowledges financial contribution from the agreement PRIN-INAF 2009 (PI: L. Sidoli).}\\
        IAPS-INAF,  via del fosso del cavaliere, 100, 00133, Rome, Italy\\
        E-mail: \email{antonella.tarana@iaps.inaf.it}}

\author{Angela Bazzano\\
        IAPS-INAF,  via del fosso del cavaliere, 100, 00133, Rome, Italy\\
        }
        
\author{Pietro Ubertini\\
        IAPS-INAF,  via del fosso del cavaliere, 100, 00133, Rome, Italy\\
        }

\abstract{We report here results on the long time behaviour of the variable source \sor\/, obtained from INTEGRAL non-continuous observations carried out between 2003 and 2009. During this period \sor\/ shows a general persistent emission along with several bright outbursts resembling those of transient X-ray sources/X-ray novae. However the source never switches into a real quiescent state, with very weak or undetectable flux level, as it normally occurs in transient sources. We compare here flux ratios and spectral state variations measured during the different outbursts and highlight the peculiar character of the 2008 outburst, for which we also propose a specific physical explanation.}

\FullConference{"An INTEGRAL view of the high-energy sky (the first 10 years)"
9th INTEGRAL Workshop and celebration of the 10th anniversary of the launch,\\
		October 15-19, 2012\\
		Bibliotheque Nationale de France, Paris, France}

\begin{document}

\section{The \sor\/ monitoring}
In the framework of a study of the Neutron Star Low Mass X-ray Binaries (NSLMXB) characteristics developed by the authors in the past years, we studied the behaviour of \sor\/, alias GRS~1724--30, a burster located in the Globular Cluster Terzan 2 \cite{grindlay80}, \cite{swank}. Its timing properties, as derived from \xte\/ results, indicated that it belongs to the Atoll class of LMXB \cite{olive}. The different flux levels detected in the past years with different observatories showed that the source is variable: while no emission above 10 keV was reported by EXOSAT \cite{psg}, hard X-ray emission with a power law spectrum with  $\Gamma$ $\sim$1.65  extending up to 100 keV \cite{barret91} was detected with SIGMA/GRANAT, making \sor\/ the first member of the NS high energy emitters class. Moreover, hard X-ray emission was again observed by \emph{BeppoSAX} and \emph{RXTE}, that was characterised by a Comptonised spectrum extending up to 200 keV, plus an additional soft component (below 3 keV) described by a blackbody emission \cite{guainazzi}.

For the present study, a first complete set of spectral states variations during the different outbursts has been measured with the same instruments providing additional information about the source behaviour over both short and long time scales.
%\section{Data analysis}
We indeed analysed  \intg\/  data  from the not-continuous, long monitoring of \sor\/ performed from October 2003 to April 2009. The used observation period consists of a total of 3650 pointings for IBIS  and 792  pointings for JEM-X, each lasting about 2000 seconds, and during which the source was, at least, in the  partially code field of view of the telescopes.% For the spectral analysis only pointings with the source within FCFOV have been used.
%%%%%%%%%%%%%%%%%%% CLUCE
\begin{figure*}[hb]
\centering
    \includegraphics[width=120mm, angle=90]{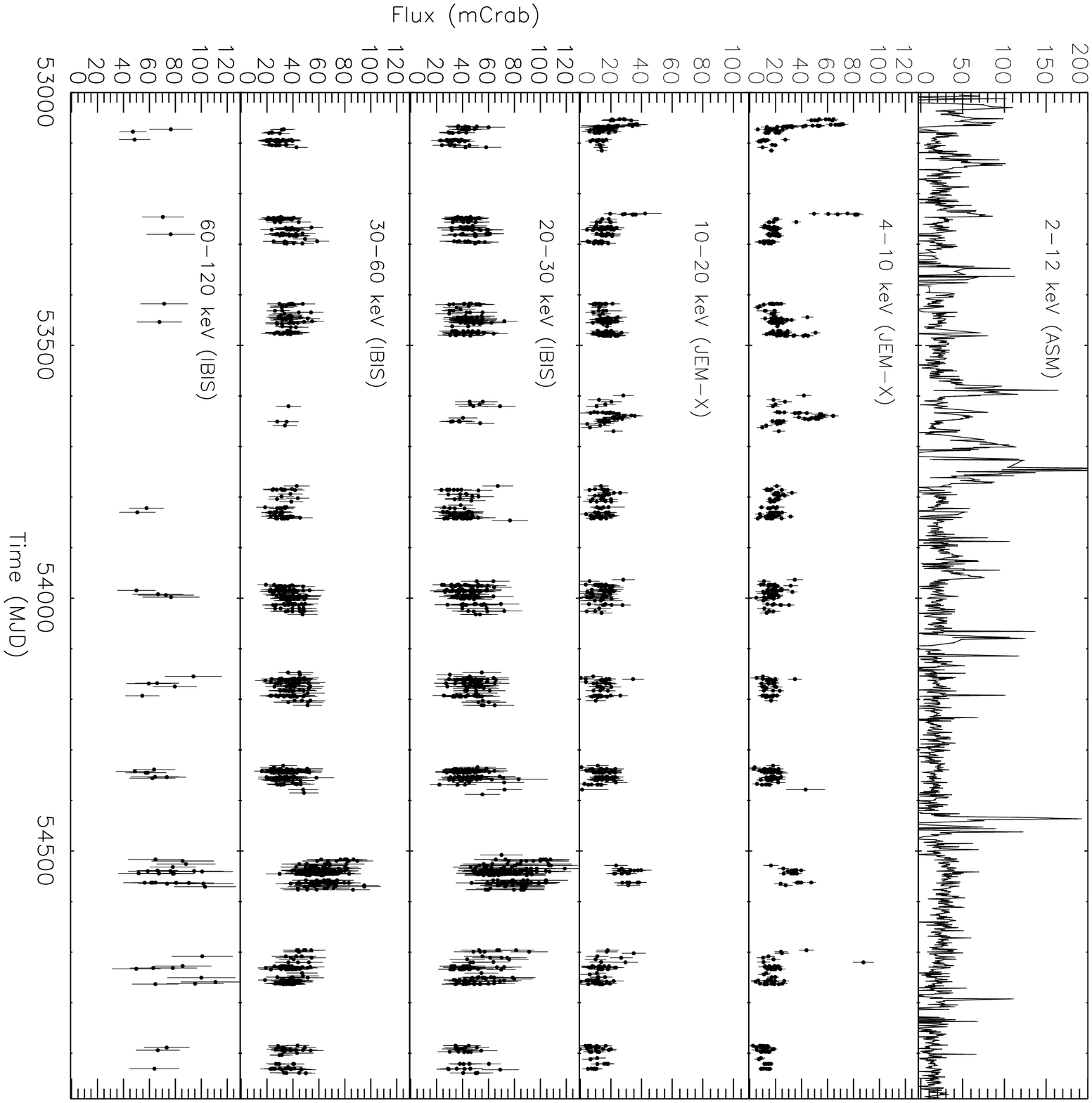}
  \caption{2003-2009 light curves of \sor\/ with \intg\/ and ASM/\xte\/.}
  \label{clucetot}
\end{figure*}
%
%%%%%%%%%%%%%%%%%%%%%%% DIAGRAMMI OUTBURSTS
\begin{figure*}
\centering
   \includegraphics[width=71mm,angle=90]{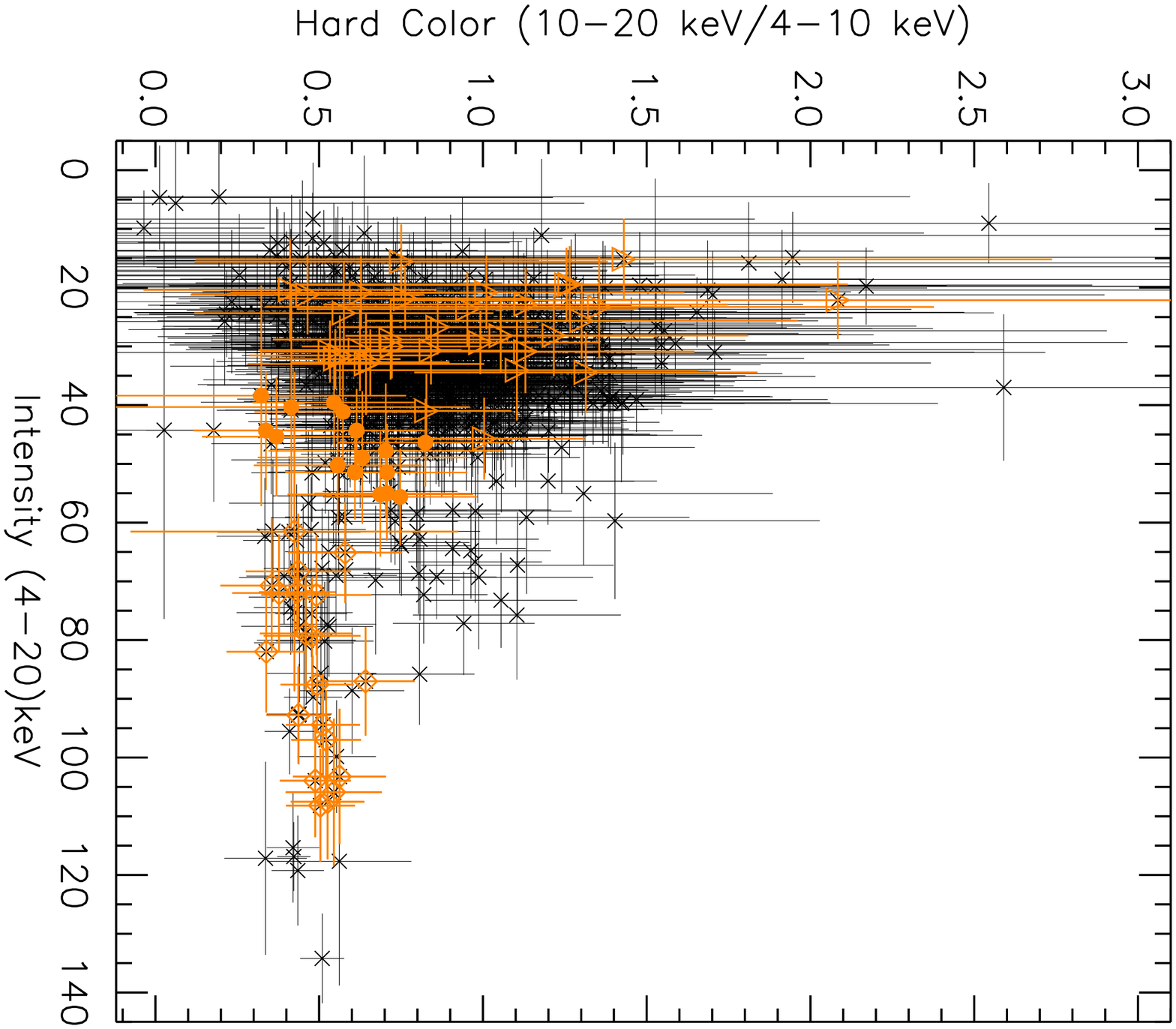}
  \vspace{1mm}
 \includegraphics[width=71mm,angle=90]{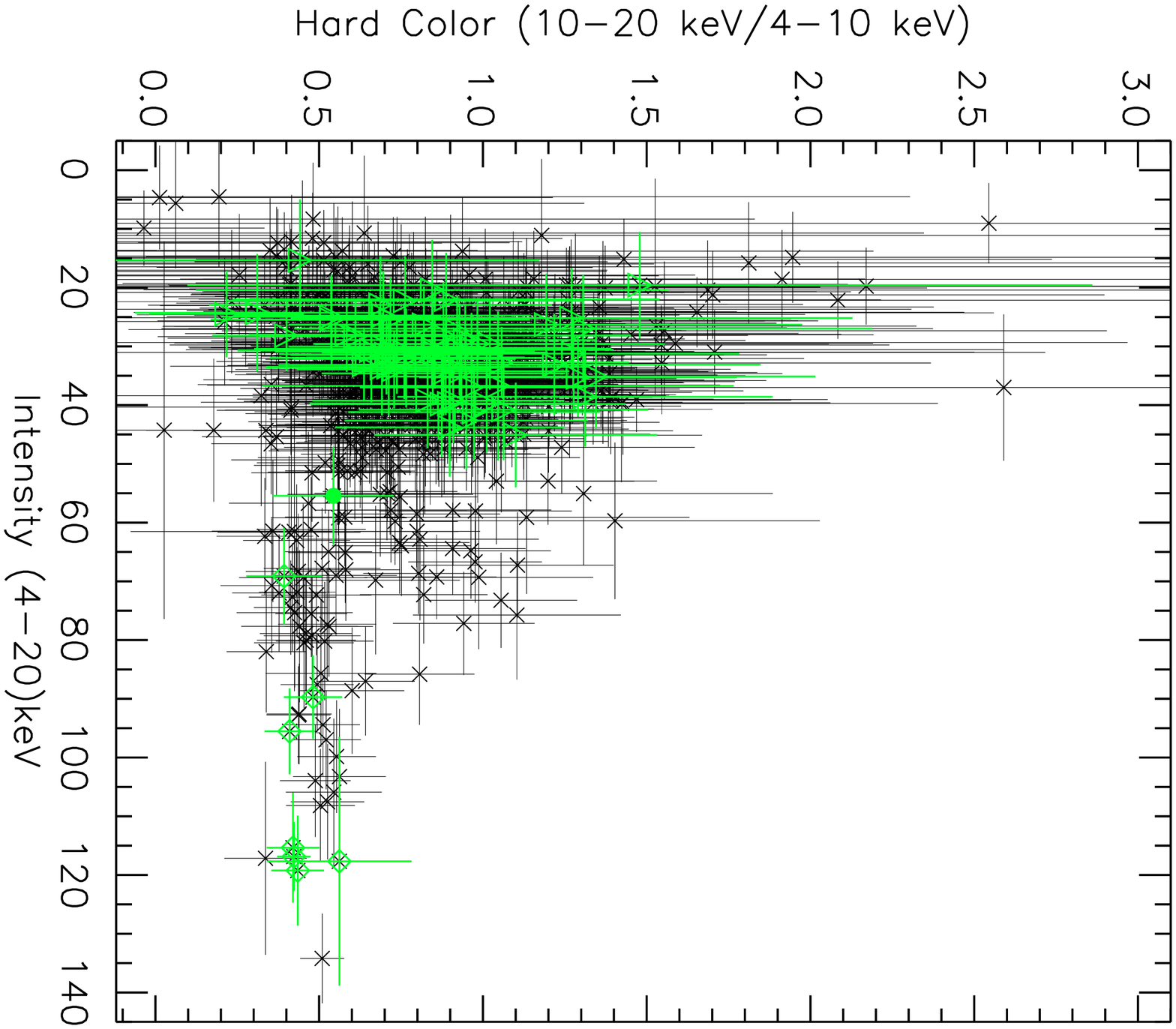}
   \vspace{1mm}
   \includegraphics[width=71mm,angle=90]{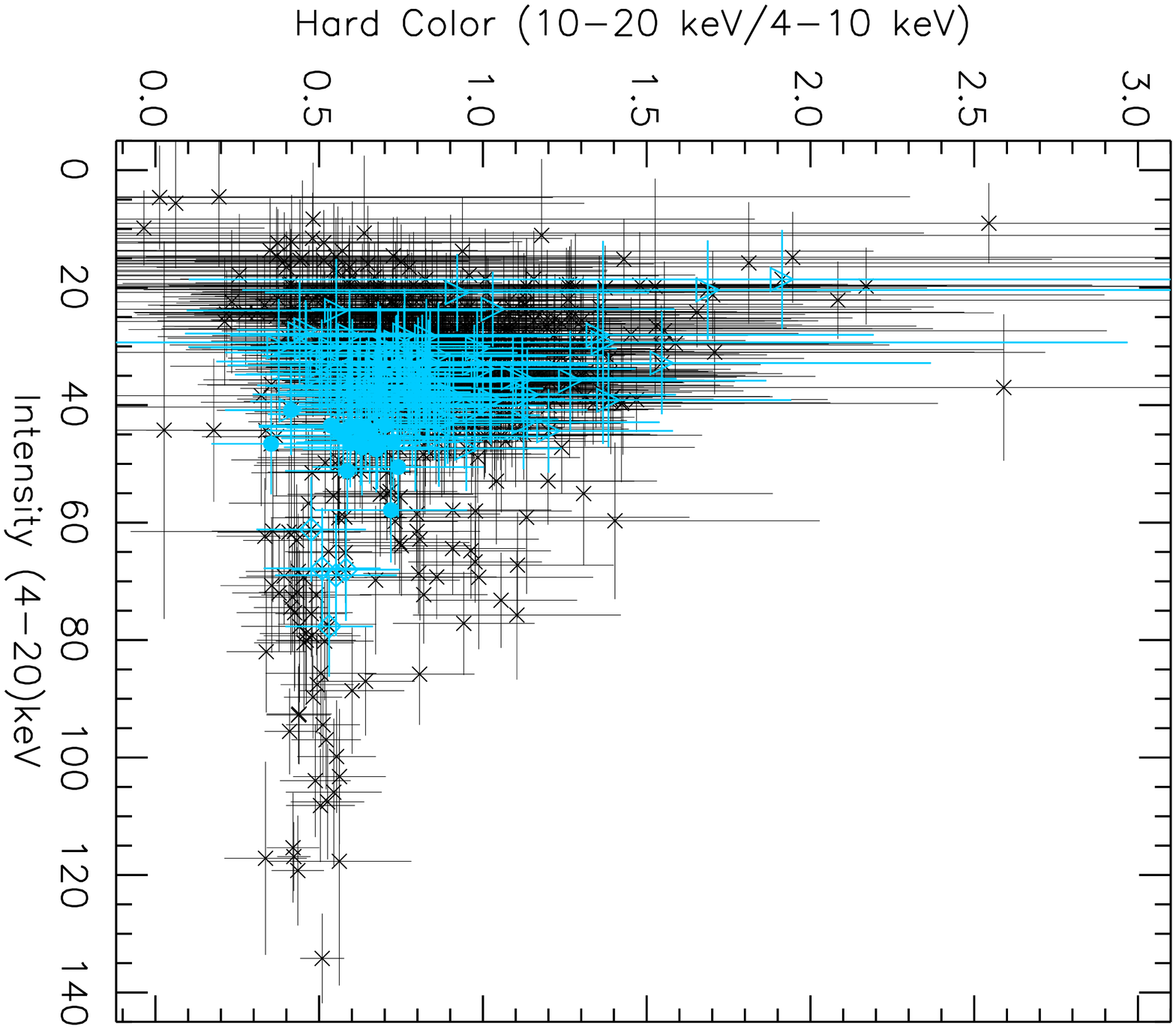}
\vspace{1mm}
   \includegraphics[width=71mm,angle=90]{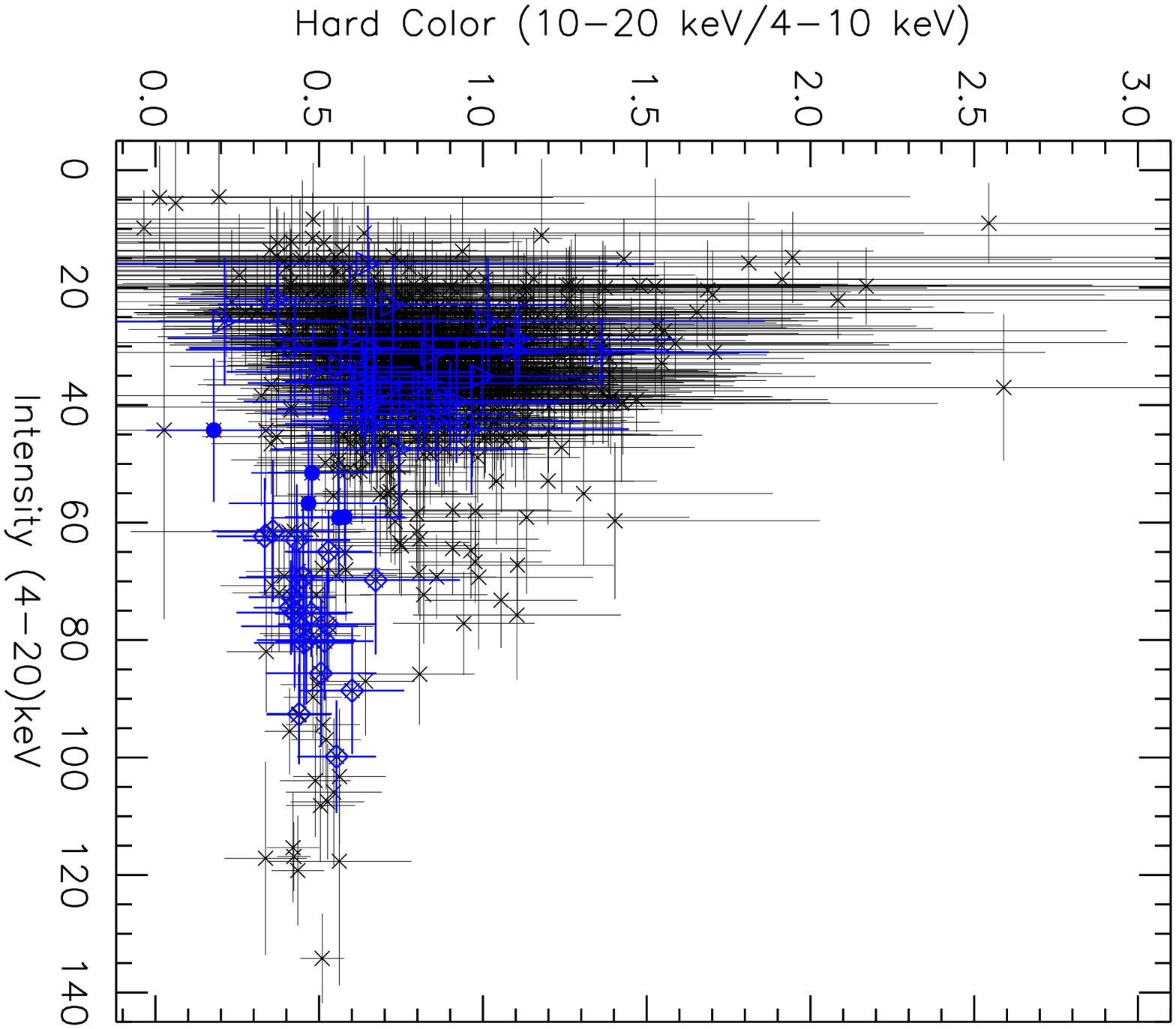}
\vspace{1mm}
   \includegraphics[width=71mm,angle=90]{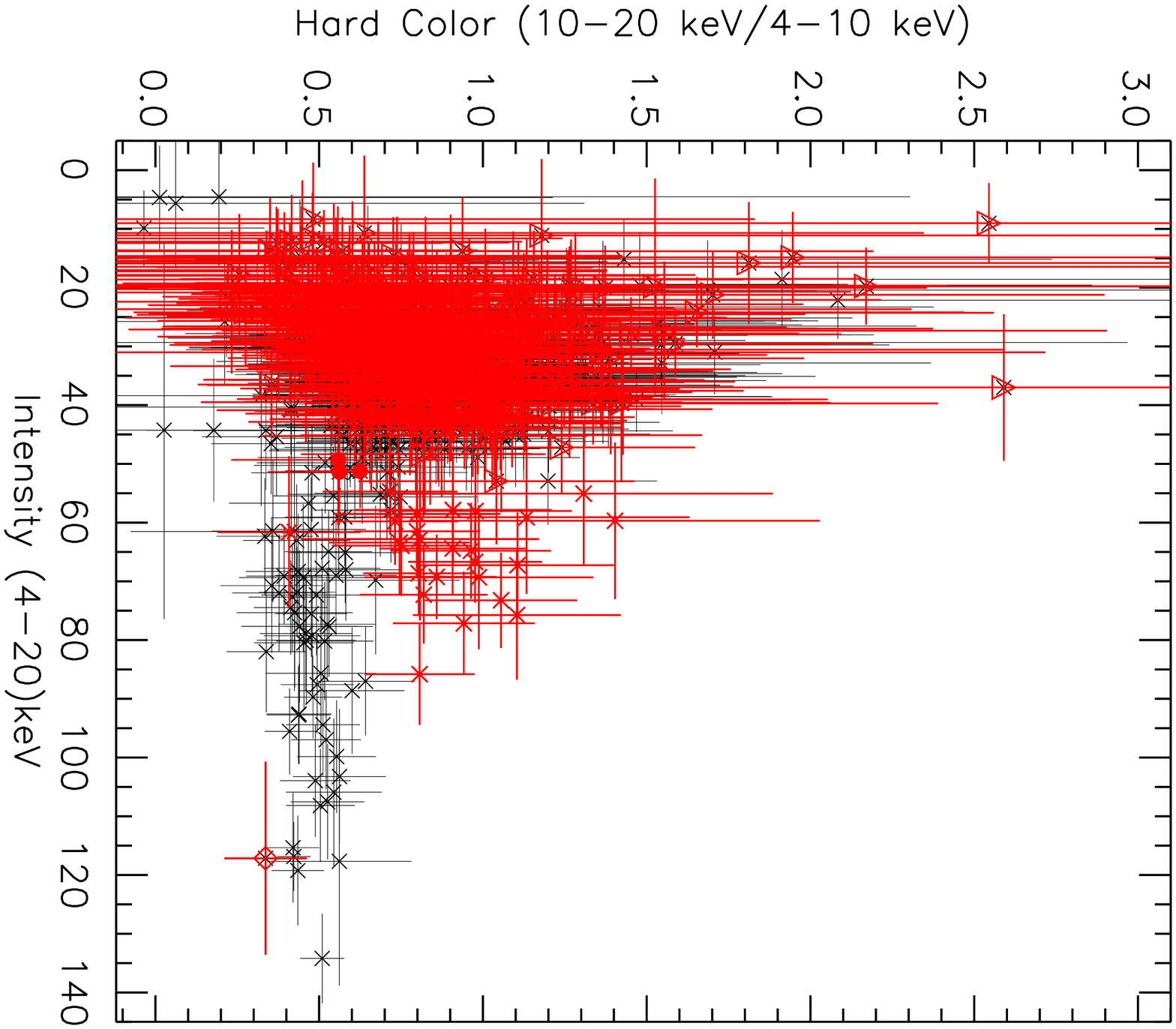}
   \vspace{1mm}
%   \includegraphics[width=60mm,angle=90]{HCI_allj_ad1-2008_ou56.eps}
% \vspace{1mm}
%    \includegraphics[width=60mm,angle=90]{HCI_allj_ad1-2008_ou2008.eps}
%\vspace{1mm}
% \includegraphics[width=60mm,angle=90]{HCI_allj_ad1-2008_ou2008_PICCO.eps}
  \caption{Color--Intensity diagrams for the different observations. The red crosses represents data from the peculiar 2008 outburst only.}%}
  \label{hci_outb}
\end{figure*}
%%%%%%%%%%%%%%%%%%%%% DIAGRAMMI
%\begin{figure} 
%\centering
%   \includegraphics[width=86.5mm,angle=90]{HCI_allij_abcB.eps}
%%  \vspace{3mm}
%  % \includegraphics[width=86.5mm,angle=90]{HCI_allj_ad1-2008.eps}
%  \caption{Total Color Intensity diagrams.}
%  \label{hci_1722}
%\end{figure}
%
%
The INTEGRAL data analysis has been performed using OSA 9.1 \cite{gol} and the hardware capability of the AVES cluster, release 1.4 \cite{memmo}, in order to menage such as large amount of data. For the spectral analysis the XSPEC software v.12.7.1 \cite{xspec} has been used. A systematic error of 2$\%$ has been added to the spectral data. The instrument constant was fixed to 1 for IBIS and kept free for JEM-X.

\subsection{Temporal variations and hardness diagrams}

Figure \ref{clucetot} shows the light curves of the whole used observation period with data from ASM/\xte\/ monitoring in the range 2-12 keV (from the web site: http://xte.mit.edu/ASM$\_$lc.html) and from JEM-X and IBIS/\intg\/ instruments in the 4-10, 10-20 keV and  20-30, 30-60 and 60-120 keV  energy bands respectively. The JEM-X and IBIS data are divided in temporal sections lasting about 3 months due the \intg\/ visibility constraints. During almost each of these temporal sections, episodes of flux increase, similar to an outburst, were detected. They were characterised by a short rise-time peak in the soft X-ray (in JEM-X and ASM data), followed by a flux increase also in the hard X-ray band (in IBIS data). During the low energy-peak the flux at high energy is lower and in some cases it is below the source detection level. These outbursts last about one month. % (first outburst: 17 Feb. 2004 -- 16 March 2004 (MJD 53058-53080). 

From the overall light curve an unique high increase of the hard X-ray emission (up to 120 mCrab for E$>$20 keV) is clearly visible near 54500 MJD (February 2008). During this episode also a "bump" in the low energy band appears as showed by JEM-X and ASM data. Before this episode (at about 54450 MJD), the source underwent an outburst in the soft X-ray band with the peak flux reaching 200 mCrab,  detected by ASM instrument only due to lack of simultaneous observations with \intg\/. These flux variations observed both in the soft and hard X-ray bands should belong to a same outburst-event characterised by a different morphology and duration (a total of about 9 months) with respect to the others.

We studied how the source moves in the color-intensity diagram during each different outburst, and the Figure \ref{hci_outb} shows the results with JEM-X data overplotted  with differents color on the complete data set of the observations. The first four outbursts (orange, green, light blue and blue chronologically) showed similar behaviour: diamond points are high intensity observations, triangles correspond to high hardness ratio values, and dots are intermediate flux observations. The later observations (after 53700 MJD)  are represented by the red data in the bottom diagram of Figure \ref{hci_outb}. In this case, the source spends most of the time in the low intensity and high hardness ratio region (red triangles data) but for the red crosses-data which correspond to a high intensity and high hardness region as appeared during the 2008 outburst described just above. The source stays in that part of the diagram only during that episode as it is clearly evident by comparing data with the other diagrams.
%%%%%%In Figure \ref{hci_1722} it is showed the hardness intensity diagram of the IBIS and JEM-X simultaneous pointings of the whole observation period. Comparing the JEM-X data only with the combined IBIS--JEM-X data no difference is noticed. We note that the source moves in the diagram only changing in flux and not in the hardness value so that it is not expected an evident spectral variation during the observations. 
%%%%%%%%%%%%%%%%%%%%%%%%% SPETTRI
\begin{figure*}
\centering
%   \includegraphics[width=60mm,angle=90]{HCI_allj_ad1-2008_ou1.eps}
%  \vspace{1mm}
% \includegraphics[width=60mm,angle=90]{HCI_allj_ad1-2008_ou2.eps}
%   \vspace{1mm}
%   \includegraphics[width=60mm,angle=90]{HCI_allj_ad1-2008_ou3.eps}
%\vspace{1mm}
\includegraphics[width=45mm,angle=-90]{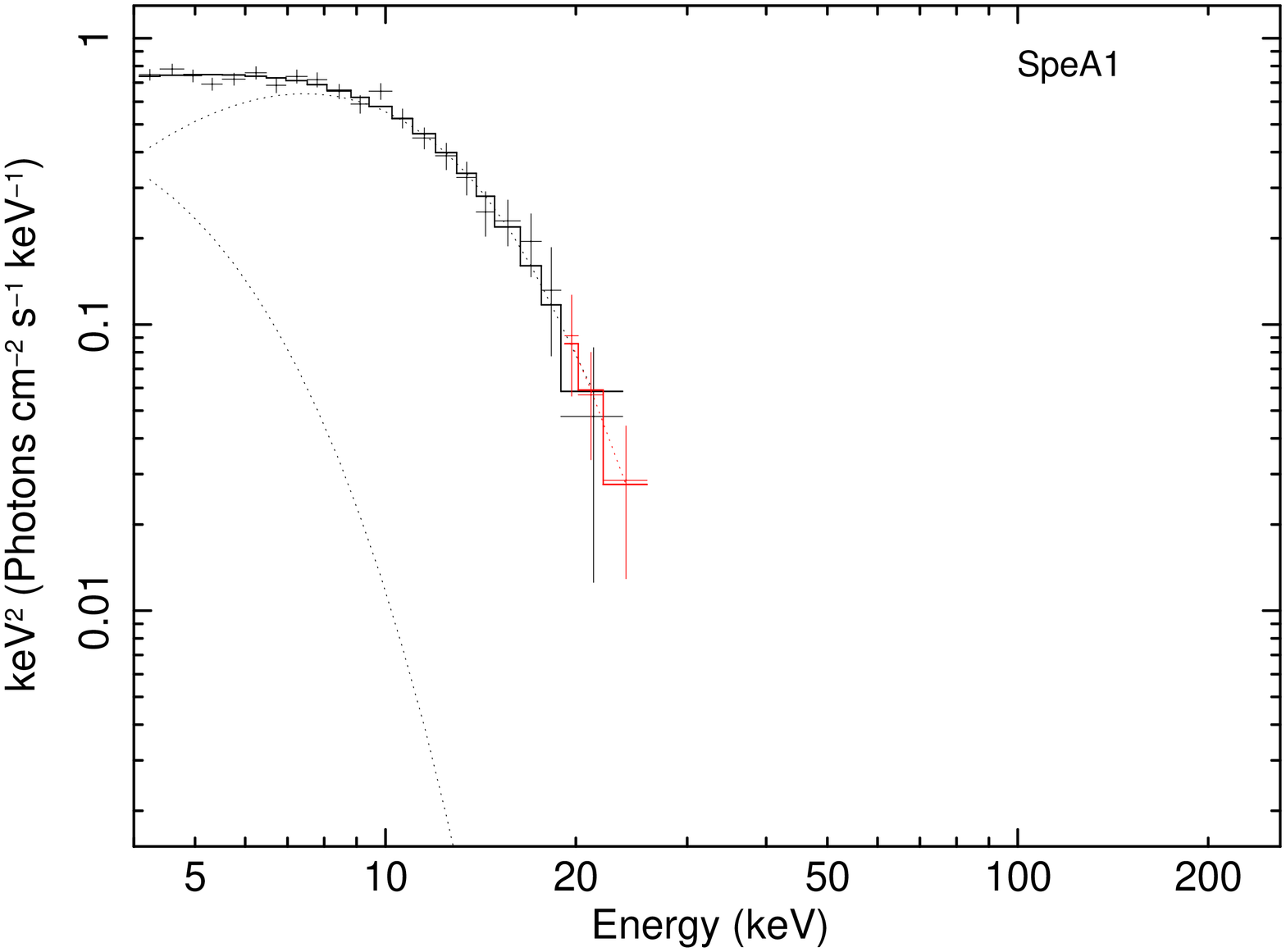}
\hspace{0.7mm}
\includegraphics[width=45mm,angle=-90]{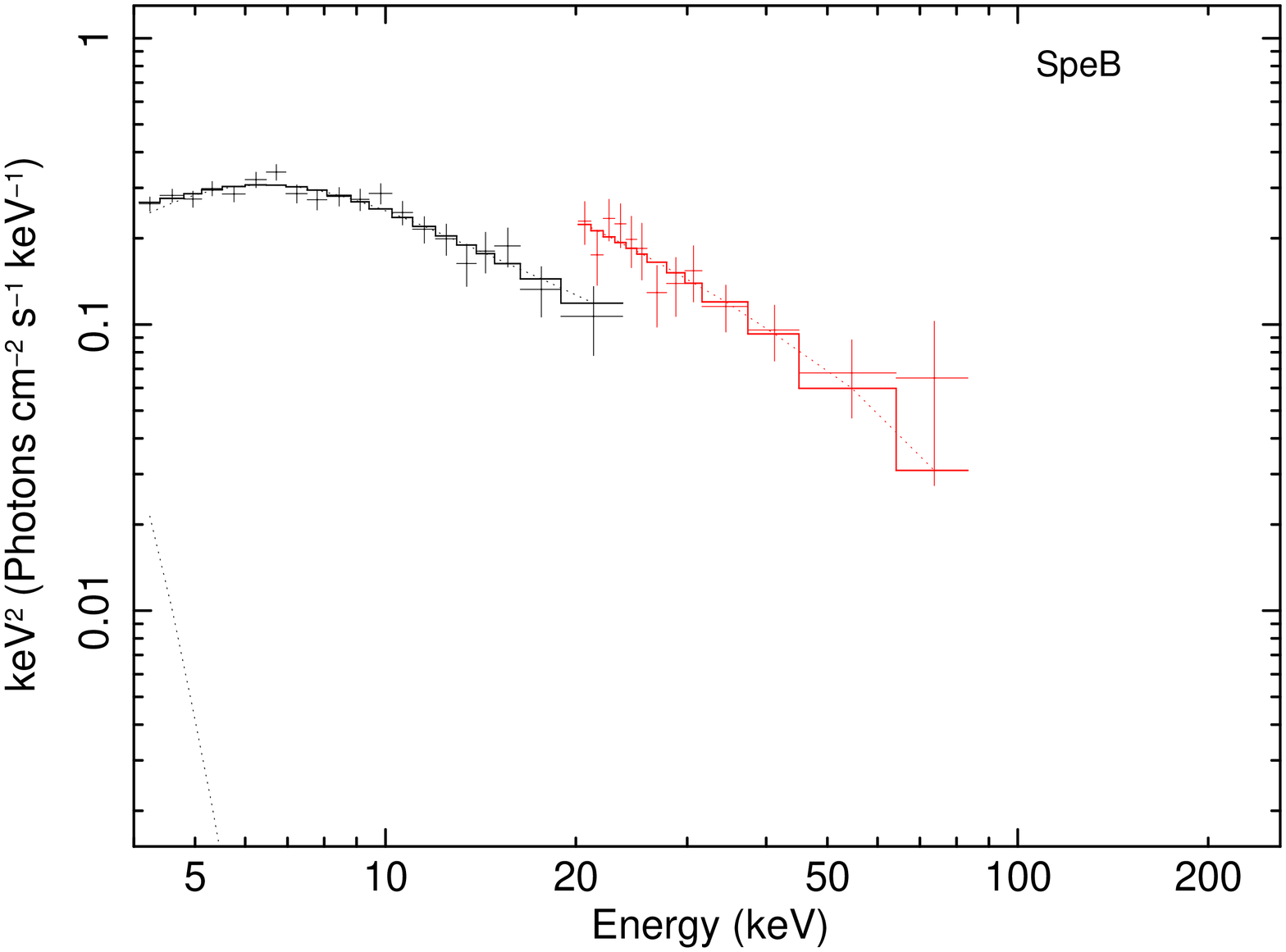}
\vspace{3mm}
\includegraphics[width=17.5mm,angle=-90]{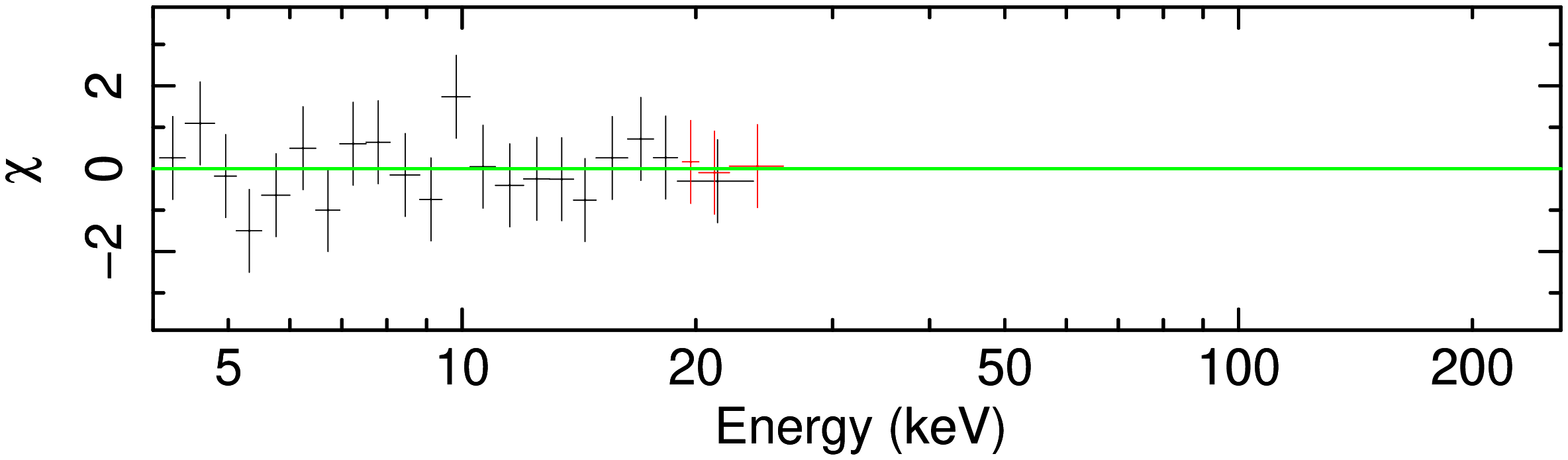}
\hspace{3mm}
\includegraphics[width=17.5mm,angle=-90]{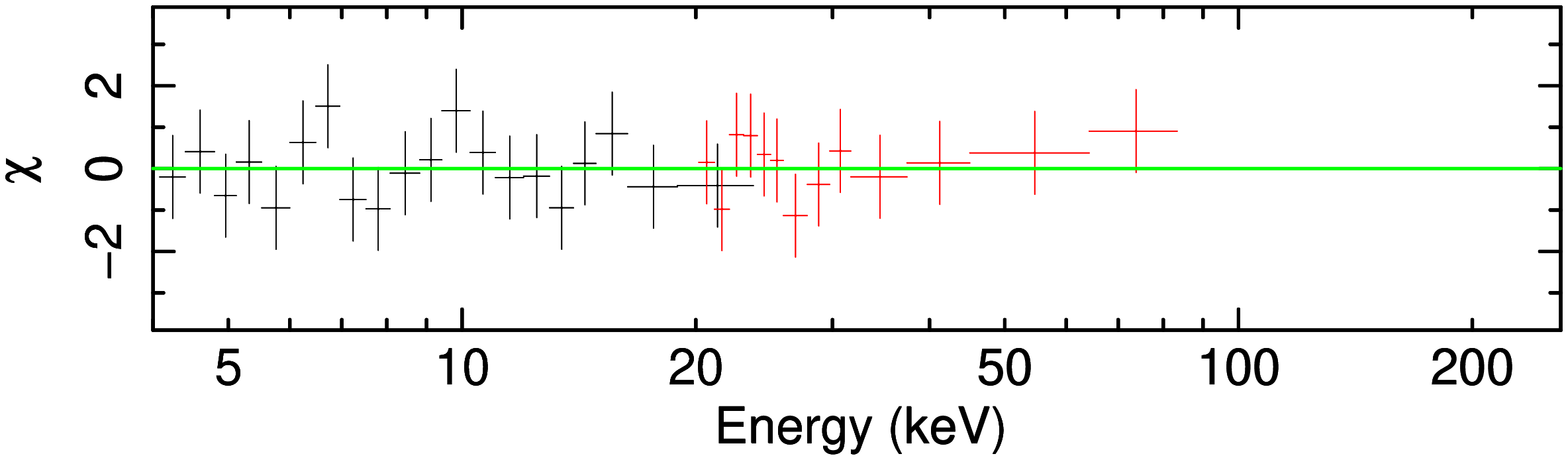}
\hspace{0.7mm}
\includegraphics[width=45mm,angle=-90]{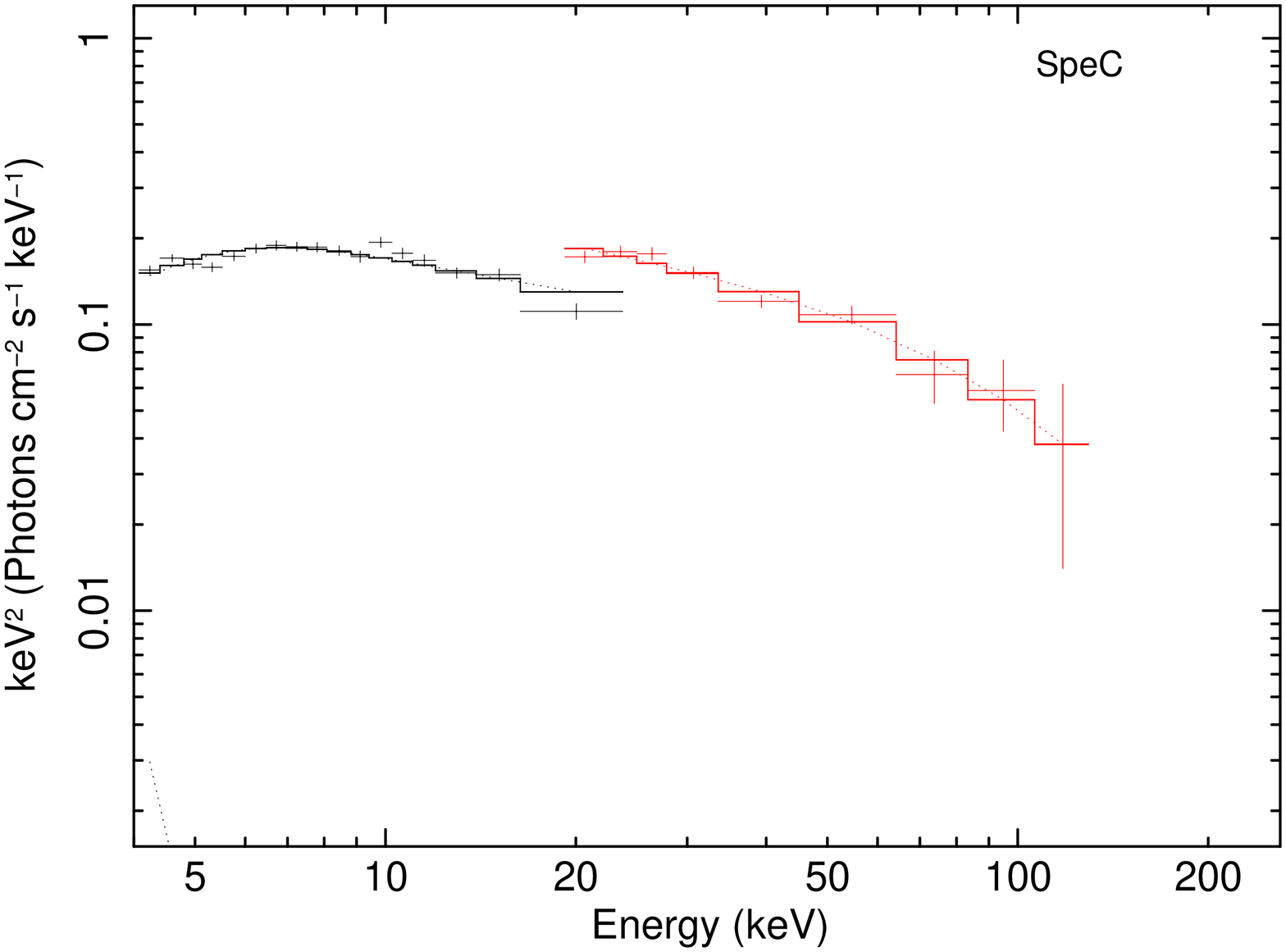}
\includegraphics[width=45mm,angle=-90]{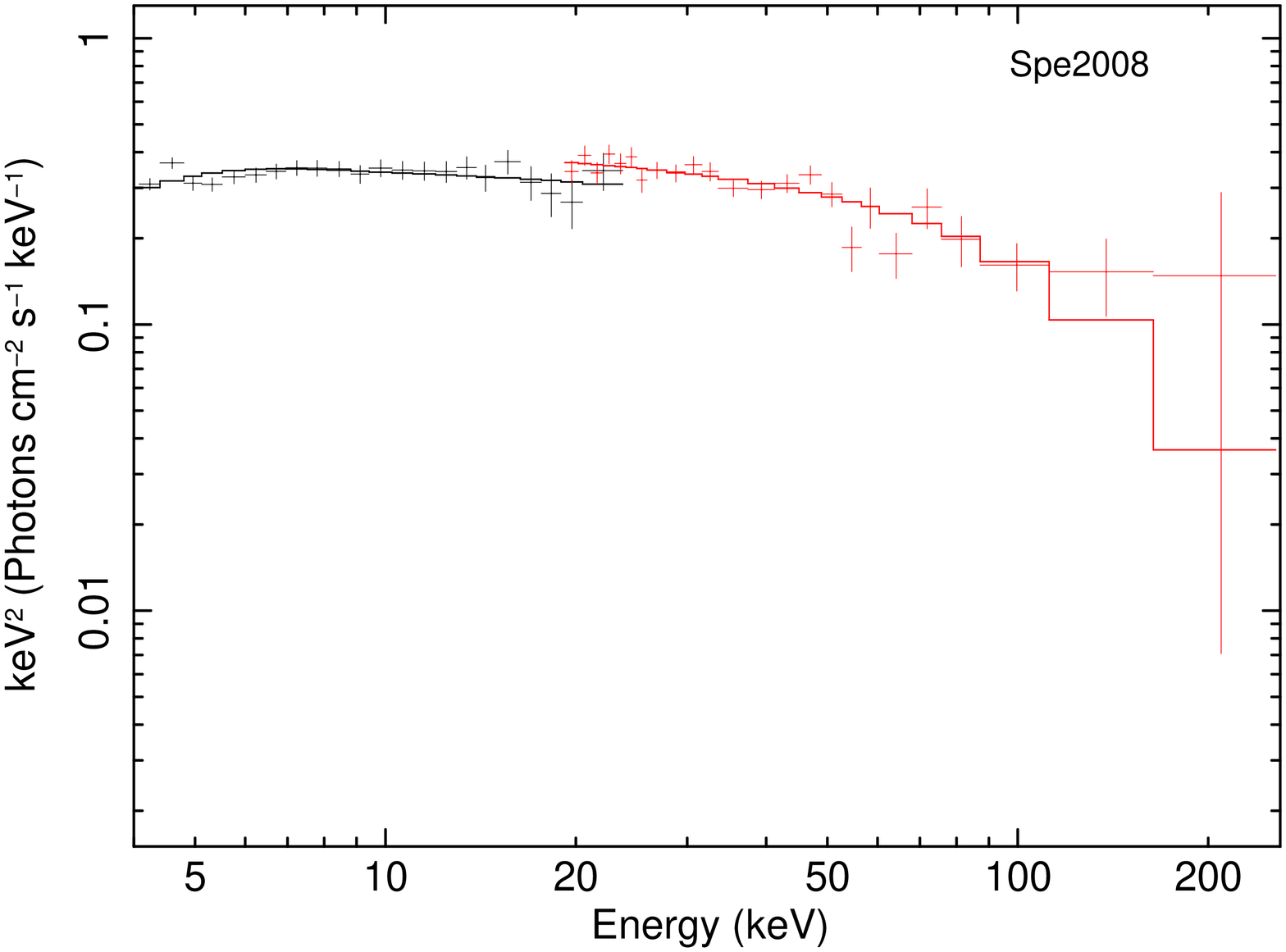}
\hspace{40mm}
\includegraphics[width=17.5mm,angle=-90]{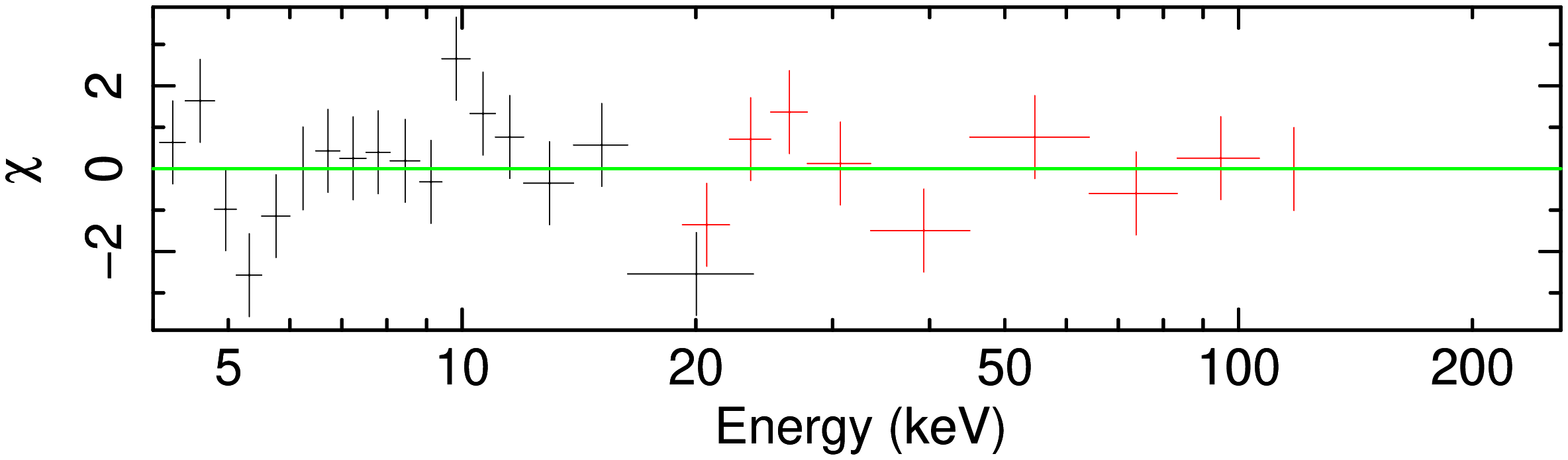}
\includegraphics[width=17.5mm,angle=-90]{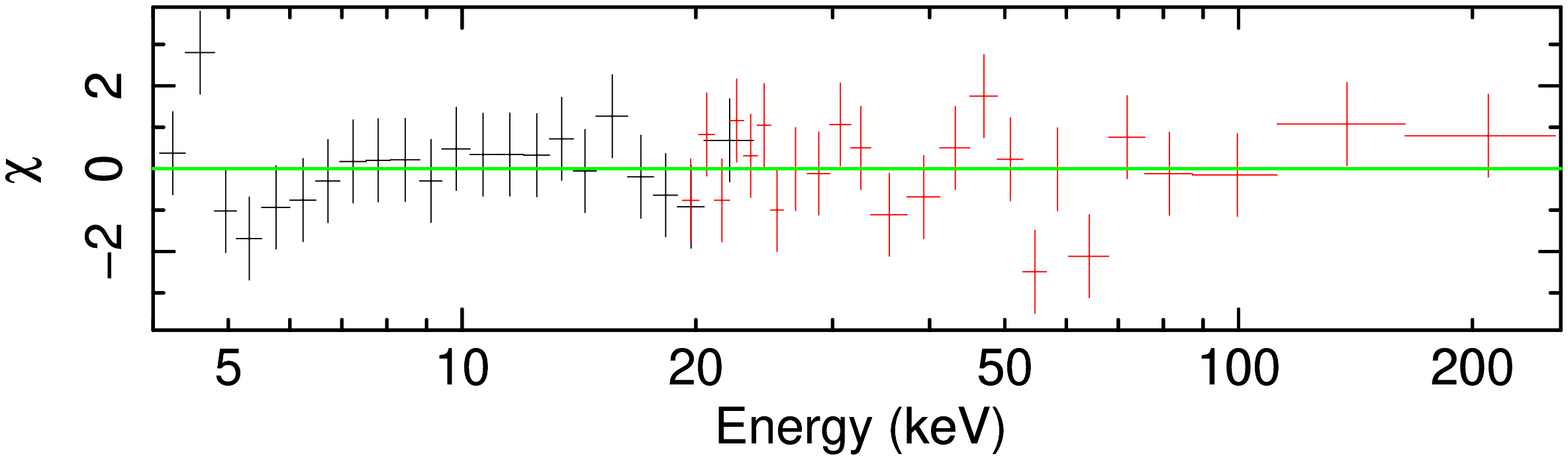}

\hspace{0.7mm}
\caption{\sor\/ spectral states with \intg\/.}
  \label{spe_1722}
\end{figure*}

\subsection{Spectral variations}

To extract the summed-in spectra with JEM-X and IBIS simultaneous pointings we collected the data in the diagrams with same symbols, even if  belonging to different outbursts. The diamond data correspond to a Soft state spectrum (named speA1); the dotted data to the Intermediate state spectrum (speB); the triangles to the Hard state spectrum (speC) and the red crosses data to the peculiar hard state of 2008 (spe2008). 
%The total exposure time is of  ks for IBIS and ks for JEM-X  for the .  
 We try to fit the spectra with different models and in Table~1 the best fit parameters obtained for the Comptonization, CompTT \cite{t}, and  black body, diskbb \cite{mitsuda}, models are reported. The spectra speC and spe2008 have slightly different  Comptonization parameter values but both correspond to a hard spectral state. It is difficult to better constrain the electron temperature of the Comptonization even adding other spectral components. %For spe2008 only upper limit of black body emission component is reported because of either  poor statistical significance or real disappearance of this component. It is difficult to constrain the electron temperature of the Comptonization also adding other spectral components. 
 %For spe2008, the black body values is reported in brackets because the fit is good even without this component  because of either poor statistical significance or real disappearance of it. 
 For spe2008, the disk black body component values are reported in Table although not constrained, because of either poor statistical significance or real disappearance of this component. The fluxes in the speA1 and speC and spe2008 correspond to 1.4$\times 10^{-9}$, 4.2$\times 10^{-10}$  and 8.6$\times 10^{-10}$ erg s$^{-1}$ cm$^{-2}$ in the 4-20 keV band. The spectra with model, data and residual are reported in Figure \ref{spe_1722}.

%%%%%%%%%%%%%%%%%%INIZIO TAB %%%%%%%%%%%%%%%%%%
\begin{table*}[h]
\vspace{-0.8cm}
\begin{minipage}{\textwidth}
\centering
\caption{Spectral fitting results for the JEM-X and IBIS spectra of 4U 1722-30.}
\label{tab_fit1722}
%\vspace{0.4cm}
\footnotesize{
\begin{tabular}{lllll}
%\colrule
%\colrule
\hline
model parameters & speA1 (diamonds) &  speB (dots) & speC (triangles) & spe2008 (red crosses)\\
%%%%\colrule
%%%\hline
%%% diskbb  & &  & & & \\
%%% \hline
%%%   $kT_{\rm in}$           & 2.49$\pm$0.22 & 2.67$\pm$0.01 &   --           & -- &      \\
%%%norm$_{\rm diskbb}$  & 3.54$\pm$1.06 &  7.66$\pm$1.61 & --  & -- &\\
%%%$\chi_{r}^2$(d.o.f)       & 0.96(28)  & 1.20(30) & $>$2 & $>$2 &\\
%%%%\colrule
%%%\hline
 %diskbb+compTT &   & & & \\
\hline
$kT_{\rm 0}$ (keV)     & 1.4(fx)\footnote{(fx) indicates parameters fixed to allow the fit to converge.}                                          & 1.2(fx)                                      &   1.1(fx)                    &  0.93(fx)       \\
$kT_{\rm e}$ (keV)                                                    & 2.42$\pm$0.50  & 19.84$\pm$17.57 & 33.60$\pm$14.52  &  30.67$\pm$8.63      \\
$\tau$                                                                         & 6.23$\pm$3.48   &  0.69$\pm$0.85     & 0.62$\pm$0.36      &  0.98$\pm$0.34     \\
norm$_{\rm CompTT}$                                              & 11.56$\pm$5.35 $\times 10^{-2}$      &1.05$\pm$1.02$\times 10^{-2}$  & 3.19$\pm$1.41$\times 10^{-3}$ &  6.29$\pm$1.72$\times 10^{-3}$\\
%\colrule
$kT_{\rm in}$ (keV)                & 0.75$\pm$0.24                & 0.35 (fx)   & 0.348$\pm$0.253 &   ($\sim$0.4)   \\
norm$_{\rm diskbb}$   & 380.27$\pm$160.73  & 8.72$\pm$2.06$\times 10^{4}$ & 0.95$\pm$7.06$\times 10^{4}$  & ($\sim$4$\times 10^{-3}$)\\
%\colrule
$\chi_{r}^2$(d.o.f) & 0.46(26)  & 0.65(42)  & 1.44(52) & 1.03(69) \\
%\colrule
\hline
%%%compTT & &  & & & \\
%%% \hline
%%% $kT_{\rm 0}$ (keV)\footnote{Fixed parameters}      & 0.44                  & 0.63$\pm$0.19 & 1   - 0.2                          &   1.1  &        \\
%%%$kT_{\rm e}$ (keV)                                                    & 2.36$\pm$0.10 & 2.59$\pm$0.21 & 23.64$\pm$22.42  -  9.10$\pm$1.45 & 39.94$\pm$18.11 &        \\
%%%$\tau$                                                                         & 7.07$\pm$0.54 & 6.77$\pm$0.83 & 0.63$\pm$0.80 -  2.36$\pm$0.33 & 0.48$\pm$0.30 &        \\
%%%norm$_{\rm CompTT}$                                              & 0.31$\pm$0.53 & 0.49$\pm$0.18 & 8.75$\pm$8.86E-03 - 0.10$\pm$4.98 &  2.77$\pm$1.31E-03 &  \\
%%%$\chi_{r}^2$(d.o.f) & 0.54(27) & 1.05(28)& 0.89(44) - 1.36(44) & 1.47(53) &\\
%%%%\colrule
%%%\hline
%%%\hline
%$F_{\rm 4-20 keV}$\footnote{The fluxes are in units of erg s$^{-1}$ cm$^{-2}$} & 3.76$\times 10^{-10}$ &  8.52$\times 10^{-10}$ \\
%$F_{\rm 20-200 keV}$  & $\times 10^{-10}$ &  $\times 10^{-10}$ \\
%$F_{\rm 60-120 keV}$ & $-$ & $-$ &1.5$\times 10^{-10}$ & $-$ & $-$ \\
%\colrule
\end{tabular}
}
\end{minipage}
\vspace{-0.66cm}
%\tablenotetext{a}{ Fixed parameters}
\end{table*}

%%%%%%%%%%%%%%%%%%FINE TAB %%%%%%%%%%%%%%%%%%
\subsection{Discussion and conclusions}

From our study, the X/hard X-ray behaviour of \sor\/ could be summarized as follow:1) We define the source in a "transient-like behaviour" during 6 years of observations: i.e. the source shows outbursts with a spectral variation similar to that observed in soft X-ray transients, but for the quiescent state presence. 2) The spectral parameters values are similar to other NSLMXBs already observed such as 4U~1820--30, 4U~1608--522 and 4U~1728--34; and during the hard states, the high energy emission is not described by a non-thermal hard power law tail component (see \cite{tarana08} for examples and details). 3) The spectral parameters variation during the outbursts is typical of other NS (and also BH) binary systems showing hysteresis cycle during outbursts: in general the temperature of the corona increases with hardening reaching value between 10-60 keV while the optical depth decreases, and the soft black body temperature decreases. 4) The source showed in 2008 a peculiar outbursts during which the flux increases both in soft and in hard X-ray band, showing a high value of the hard flux ($>$ 20 keV) never reached by the other hard states observed during our monitoring. The simplest hypothesis that can explain the peculiar outburst is that the source reached a regime of particular high accretion rate due to accretion disk instability episodes. However we should also take into account the changes of X-ray flux connected to super orbital variation. In fact, for this source, it was found a 12 yr quasi periodic modulation explained as variations of the donor star magnetic field during its solar-like cycle \cite{kot}. This turns out as orbital period variations and consequently as Roche radius and accretion rate changes. The last flux increase in that modulation falls on just near the 2008 outburst observed by us, showing that also this contribution for the accretion rate increase could be present. Further observations are necessary to understand the validity of the super orbital modulation - peculiar outbursts connection.


\vspace{-0.5cm}
\begin{thebibliography}{99}
\vspace{-0.3cm}
\footnotesize{
\bibitem{grindlay80} Grindlay, J. E., Marshall, H. L., Hertz, P., et al. ApJ 240, L21, 1980
\bibitem{swank} Swank, J.H., Becker, R. H., Boldt, E. A., et al.  ApJL 212 L73, 1977
\bibitem{olive} Olive, J.F., Barret, D., Boirin, L. et al.  A\&A 333, 942, 1998 
\bibitem{psg} Parmar A.N., Stella, L. \& Giommi, P., A\&A 222, 96, 1989
\bibitem{barret91} Barret, D., Olive, J.F., Boirin, L. et al. A\&A 357, L41, 1991
\bibitem{guainazzi} Guainazzi M., et al. A\&A 339, 802, 1998
\bibitem{gol} Goldwurm A., David P., Foschini L. et al  A\&A, 411, L223, 2003
\bibitem{memmo} Federici, M. \& Martino B.  \pos{PoS(extremesky2009)092},  2009
\bibitem{xspec} Arnauld K. A., in Jacoby G.H., Bames J., eds, ASP Conf. Ser. Vol 101, Astronomical data analysis software and systems V. Astron. Soc. Pa., San Francisco, p.17
\bibitem{t} Titarchuk, L.  ApJ 434, 570, 1994
\bibitem{mitsuda} Mitsuda, K. et al., PASJ, 36, 741, 1984
\bibitem{tarana08} Tarana A., Bazzano A. and Ubertini P.,  \pos{PoS(Integral08)044}, 2008
%\bibitem{eme} Emelyanov, A. N.; Revnivtsev, M. G.; Aref'ev, V. A.; Sunyaev, R. A., AstL, 28, 12, 2002
\bibitem{kot} Kotze, M. M.; Charles, P. A., MNRAS, 402L, 16, 2010.
%
%
%\bibitem{xspec} Arnauld K. A., in Jacoby G.H., Bames J., eds, ASP Conf. Ser. Vol 101, Astronomical data analysis software and systems V. Astron. Soc. Pa., San Francisco, p.17
%\bibitem{gol} Goldwurm A., David P., Foschini L. et al  A\&A, 411, L223, 2003
%\bibitem{guainazzi} Guainazzi M., et al. A\&A 339, 802, 1998
%\bibitem{grindlay80} Grindlay, J. E., Marshall, H. L., Hertz, P., et al. ApJ 240, L21, 1980
%\bibitem{swank} Swank, J.H., Becker, R. H., Boldt, E. A., et al.  ApJL 212 L73, 1977
%\bibitem{olive} Olive, J.F., Barret, D., Boirin, L. et al.  A\&A 333, 942, 1998 
%\bibitem{barret91} Barret, D., Olive, J.F., Boirin, L. et al. A\&A 357, L41, 1991
%\bibitem{psg} Parmar A.N., Stella, L. \& Giommi, P., A\&A 222, 96, 1989
%\bibitem{memmo} Federici, M. \& Martino B.  \pos{PoS(extremesky2009)092},  2009
%\bibitem{mitsuda} Mitsuda, K. et al., PASJ, 36, 741, 1984
%\bibitem{t} Titarchuk, L.  ApJ 434, 570, 1994
%%\bibitem{Kuu} Kuulkers, E., den Hartog, P.~R., in't Zand  J.~J.~M. et al.  2003, A\&A, 399, 663 
%% ,Verbunt, F.~W.~M., Harris, W.~E., \& Cocchi, M., 2003, A\&A, 399, 663 
%\bibitem{tarana08} Tarana A., Bazzano A. and Ubertini P.,  \pos{PoS(Integral08)044}, 2008
%%\bibitem{eme} Emelyanov, A. N.; Revnivtsev, M. G.; Aref'ev, V. A.; Sunyaev, R. A., AstL, 28, 12, 2002
%\bibitem{kot} Kotze, M. M.; Charles, P. A., MNRAS, 402L, 16, 2010.
}
\end{thebibliography}
\end{document}